\newcommand{\Eq}[1]{Eq.~({\protect\ref{#1}})}

\newcommand{\Ref}[1]{Ref.\protect\cite{#1}}

\newcommand{\Sec}[1]{Sect.~\protect\ref{#1}}
\newcommand{\Fig}[1]{Fig.~\protect\ref{#1}}

\newlength{\Tatescale}
\newcommand{\agt}{\mbox{ \raisebox{0.4ex}{$>$}\hspace{-0.8em}\raisebox{-0.9ex}{$\sim$} }}
\newcommand{\alt}{\mbox{ \raisebox{0.4ex}{$<$}\hspace{-0.8em}\raisebox{-0.9ex}{$\sim$} }}

\newlength{\figwidth}
\documentclass[epj]{svjour}
\usepackage{graphics}
\begin{document}
\title{A remark on the large  difference between the glueball mass and
\boldmath{$T_c$} in quenched QCD}
\author{Noriyoshi ISHII\inst{1} \and Hideo SUGANUMA\inst{2}
}                     
\institute{
Radiation Lab.,
The Institute of Physical and Chemical Research (RIKEN),
2-1 Hirosawa, Wako, Saitama 351-0198, Japan\\
\email{ishii@rarfaxp.riken.go.jp}
\and
Faculty of Science, Tokyo Institute of Technology,
2-12-1 Ohokayama, Meguro, Tokyo 152-8551, Japan\\
\email{suganuma@th.phys.titech.ac.jp}
}
\date{Received: date / Revised version: date}
\abstract{
The lattice  QCD studies indicate  that the critical  temperature $T_c
\simeq 260-280$ MeV of  the deconfinement phase transition in quenched
QCD  is  considerably  smaller  than the  lowest-lying  glueball  mass
$m_{\rm G} \simeq 1500-1700$ MeV, i.e., $ T_c \ll m_{\rm G}$.
As a consequence  of this large difference, the  thermal excitation of
the glueball  in the confinement  phase is strongly suppressed  by the
statistical factor as $e^{-m_{\rm G}/T_c} \simeq 0.00207$ even near $T
\simeq T_c$.
We consider  its physical implication, and argue  the abnormal feature
of  the  deconfinement  phase  transition  in quenched  QCD  from  the
statistical viewpoint.
To appreciate this,  we demonstrate a statistical argument  of the QCD
phase transition using the recent lattice QCD data.
From the phenomenological relation  among $T_c$ and the glueball mass,
the deconfinement  transition is found  to take place in  quenched QCD
before a reasonable amount of glueballs is thermally excited.
In this way, quenched QCD reveals  a question ``what is the trigger of
the deconfinement phase transition ?''
} 
\authorrunning{N.~Ishii, H.~Suganuma}
\maketitle

\section{Introduction}
\label{sec.intro}

The quark-gluon-plasma (QGP) is one of the most interesting targets in
the  finite-temperature quark-hadron  physics  \cite{qcd}.  Currently,
the  QGP creation experiment  is being  performed in  the relativistic
heavy  ion   collider  (RHIC)  project  at   the  Brookhaven  National
Laboratory   (BNL),   and   much   progress   in   understanding   the
finite-temperature QCD is desired.
Historically, the instability of the  hadron phase was first argued by
Hagedorn \cite{hagedorn}  before the discovery of QCD.  He pointed out
the possibility of a phase  transition at finite temperature, based on
the string or the flux-tube picture of hadrons\cite{hagedorn,patel}.
After  QCD was  established as  the fundamental  theory of  the strong
interaction, this phase transition was recognized as the deconfinement
phase  transition  to the  QGP  phase,  where  quarks and  gluons  are
liberated with the restored chiral symmetry.
The QCD phase transition  has been studied using various QCD-motivated
effective models such as  the linear $\sigma$ model\cite{kapusta}, the
Nambu-Jona-Lasinio      model\cite{hatsuda-kunihiro},     the     dual
Ginzburg-Landau theory\cite{ichie} and so on.

In  order   to  study  nonperturbative  features  of   the  QCD  phase
transition,  the  lattice QCD  Monte  Carlo  calculation  serves as  a
powerful tool directly based on QCD.
It has  been already extensively used  to study the nature  of the QCD
phase transition.  At the  quenched level, SU(3) lattice QCD indicates
the existence  of the deconfinement  phase transition of a  weak first
order at $T_c \simeq 260-280$  MeV \cite{karsch2}.  On the other hand, 
in the  presence of  dynamical quarks, it  indicates the  chiral phase
transition at  $T_c = 173(3)$ MeV  for $N_f=2$ and $T_c  = 154(8)$ MeV
for $N_f = 3$ in the chiral limit \cite{karsch}.

For the  comparison with the experimental  data in the  real world, it
would  be desirable  to investigate  full QCD  with  dynamical quarks.
However, there  are a  number of underlying  nonperturbative features,
which are shared  in common by both full QCD and  quenched QCD such as
color  confinement and  instanton phenomena.   In order  to understand
such nonperturbative features of QCD, quenched QCD provides us with an
idealized environment  to focus on  the essence of the  problem itself
without involving inessential technical complexities.

The aim  of this  paper is to  point out  the abnormal feature  of the
large difference  between the lowest-lying glueball mass  $m_{\rm G} =
1500-1700$ MeV  and the  critical temperature $T_c  = 260-280$  MeV in
quenched QCD.
In fact, due to  this large difference,  the thermal exciation  of the 
glueball  is strongly  suppressed  by  a small statistical  factor
$e^{-m_{\rm G}/T_c} \simeq 0.00207$ even near $T\simeq T_c$.
In order to appreciate this point, we
demonstrate a statistical argument on the QCD phase transition 
using the closed packing model with the bag-model picture of hadrons.
The  closed packing model  has  been  often  used  for  the  phenomenological
understanding of the full QCD phase transition near $T_c$ from below.
In the confinement phase, only color-singlet states such as hadrons 
can contribute to the Boltzmann sum in  the partition function, 
and therefore it is desirable to understand the QCD phase transition 
in terms of the hadronic degrees of freedom below $T_c$.

The contents are organized as follows. In \Sec{sec.closed.packing}, we
give a brief review of the closed packing model, i.e., the statistical
treatment of  the QCD phase  transition with the bag-model  picture of
hadrons.   We  derive a  phenomenological  relation  among the  hadron
masses,   the  hadron   sizes  and   the  critical   temperature.
In \Sec{sec.full.qcd}, 
we apply the statistical approach to full QCD and compare it with the recent lattice data of full QCD. 
In  \Sec{sec.quench.qcd}, we apply the statistical approach to  quenched  QCD, 
and demonstrate an abnormal feature of the quenched
QCD phase  transition. In  \Sec{sec.discussion}, after the  summary of
the results,  we attempt to clarify  the essence of  this problem, and
discuss an  abnormal nature of  the deconfinement phase  transition in
quenched QCD.

\section{A review of the statistical approach}
\label{sec.closed.packing}

In this section,  we give a brief review of  the closed packing model,
i.e., the statistical  treatment of the QCD phase  transition based on
the bag-model picture of hadrons.
It provides us with a simple but useful insight into the QCD phase 
transition from below $T_c$ in terms of hadronic properties.
We derive a phenomenological relation between the critical temperature
$T_c$ and the properties of the hadrons, i.e., the mass and the size.

In the bag-model picture, quarks and gluons are assumed to be confined
inside  a  spherical bag.   Color  confinement  is  simply taken  into
account through the bag-like intrinsic structure of hadrons \cite{jaffe}.
At  low temperature,  only a  small number  of such  bags  are thermally
excited, and the thermodynamic properties of the system are described in
terms of these spatially isolated bags.
With the  increasing temperature, the number of  the thermally excited
bags increases.   Gradually, these bags begin to  overlap one another,
and  they  finally  cover  the   whole  space  region  at  a  critical
temperature $T_c$ in this picture.
Above $T_c$,  as a result of overlapping bags, the whole
space   is  filled  with   liberated  quarks   and  gluons,   and  the
thermodynamic properties of  the system is now governed  by quarks and
gluons.
In this  way, the QCD  phase transition is  described in terms  of the
overlaps of  the thermally  excited bags.

\begin{figure}
\resizebox{0.48\textwidth}{!}{
  \includegraphics{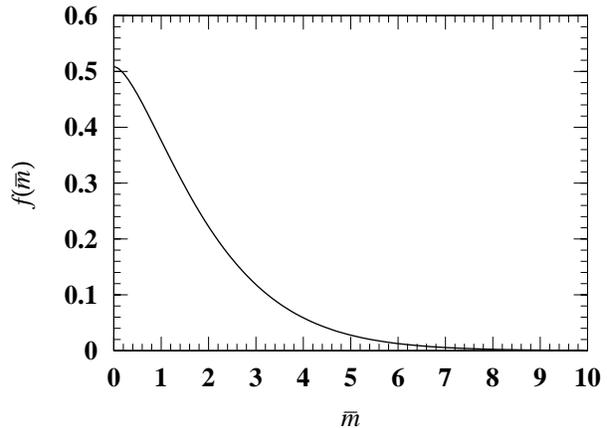}
}
\caption{The characteristic function $f(\bar m)$ on the thermal boson 
in \Eq{fmt} plotted  against $\bar m=m/T$.}
\label{f-m-t}
\end{figure}

For the quantitative argument, we  define  the  spatial occupation  ratio  $r_V(T)$  at
temperature $T$  to be the  ratio of the  total volume of  the spatial
regions inside  the thermally  excited bags to  the volume $V$  of the
whole space  region.  In the closed  packing picture of  the QCD phase
transition, $r_V(T)$ plays the key role, which is estimated as
\begin{eqnarray}
	r_V(T)
&=&
	\frac1V \sum_n{4\pi\over  3} R_n^3 \cdot  \lambda_n N_n(T) 
\label{rv}
\\\nonumber
&=&
	\sum_n 
	\lambda_n 
	R_n^3
	{4\pi \over 3}
	\int {d^3k\over(2\pi)^3}
	{1\over e^{\sqrt{m_n^2 + \vec k^2}/T} - 1} 
\\\nonumber
&=&	\sum_n 
	\lambda_n 
	R_n^3 T^3 f(m_n/T),
\end{eqnarray}
where  $N_n(T)$,  $\lambda_n$,  $m_n$  and  $R_n$ are  the  number  at
temperature $T$,  the degeneracy, the mass  and the bag  radius of the
$n$-th elementary  excitation, respectively. 
Here, $f(\bar m)$ is defined by
\begin{eqnarray}
	f(\bar m)
&\equiv&
	{4\pi \over 3}
	\int {d^3\bar k\over(2\pi)^3}
	{1\over e^{\sqrt{\bar m^2 + \bar k^2}} - 1},
\label{fmt}
\end{eqnarray}
and  its functional form  is plotted  against $\bar  m \equiv  m/T$ in
\Fig{f-m-t}.   Note  that $f(m/T)$  is  a  characteristic function  to
describe the  thermal contribution of the  boson with the  mass $m$ at
the temperature  $T$ \cite{ST91}.  For  $m \gg T$,  $f(m/T)$ decreases
exponentially with $m/T$, and  the thermal contribution is expected to
become negligible.

In the closed packing picture, the phase transition is assumed to 
takes place,  when the thermally  excited bags almost cover  the whole
space region.   Hence, the critical temperature $T_c$  is estimated by solving
\begin{equation}
	r_V(T_c) = 1.
\label{critical.temperature}
\end{equation}

In  the closed packing  picture, an  essential role  is played  by the
color confinement,  which is a  peculiar phenomenon in QCD  making the
underlying quark and gluon structure hidden inside the hadron.
This feature  is quite different  from atomic system.   (For instance,
the closed  packing picture of  phase transition cannot be  applied to
the Coulombic  system such as  the ionization transition of  an atomic
gas,   where  the   bag-model  description   of  the   bound-state  is
inappropriate.)
The readers might  feel that the closed packing  argument is too naive
for the complicated QCD phase transition.
In particular,  the ideal-gas statistical treatment  should be applied
only to the dilute system in a strict sense.
Nevertheless,  the  closed  packing  approach provides  a  simple  and
physical insight into the complicated QCD phase transition in terms of
the thermal excitations of hadrons.
Hence,  it would  be natural  to attempt  understanding the  QCD phase
transition in the closed packing picture as a first step.

\section{A statistical approach to the full-QCD phase transition}
\label{sec.full.qcd}
First, we apply the statistical approach with the closed packing model
to the  full QCD  phase transition, and  compare the results  with the
recent lattice QCD data.
In full QCD, the lightest physical excitation is the pion, and all the
other hadrons are rather heavy as  $m \gg m_\pi$, $T_c$.  In fact, the
pion  is   considered  to  play   the  key  role  in   describing  the
thermodynamic properties of full QCD below $T_c$ from the viewpoint of
the statistical physics.
Hence, in  most cases,  only the pionic  degrees of freedom  are taken
into account in the hadron phase in the argument of the full-QCD phase
transition.
By using the isospin degeneracy $\lambda_{\pi} = 3$, the mass $m_{\pi}
=  140$  MeV   and  the  radius  $R_{\pi}  \simeq   1$  fm,  we  solve
\Eq{critical.temperature}  with $r_V(T)=3 R_{\pi}^3  T^3 f(m_{\pi}/T)$
to estimate the critical temperature as $T_c \simeq 183$ MeV.
Considering its closeness to the full lattice QCD result with $N_f=2$,
i.e., $T_c \simeq  170$ MeV, the statistical approach  to the full-QCD
phase transition seems to be rather good.

We now  consider the $m_{\pi}$-dependence of  the critical temperature
$T_c$. Note  that, in  the actual lattice  QCD calculations,  the pion
mass is taken  to be still rather heavy as $m_{\pi}  \agt 400$ MeV for
the technical reasons.
From these data, the critical temperature $T_c$ in the chiral limit is
obtained using the chiral extrapolation.
\begin{figure}
\resizebox{0.48\textwidth}{!}{
  \includegraphics{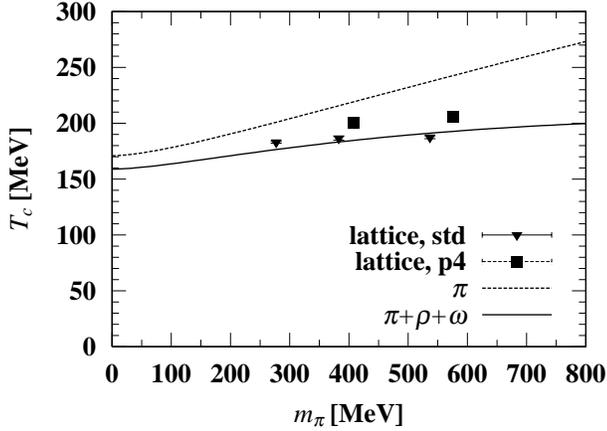}
}
\caption{ The critical temperature $T_c$ plotted against the pion mass
$m_{\pi}$ in  the $N_f=2$ case.   The dashed curve denotes  the result
retaining only  the contribution from pions.  The  solid curve denotes
the  result  including also  contributions  from  $\rho$ and  $\omega$
mesons.    The  triangle   denotes   the  lattice   data  taken   from
\Ref{luetgemeier}  obtained with standard  (std) fermion  action.  The
square denotes the lattice  data taken from \Ref{karsch} obtained with
the improved staggered (p4) fermion action. }
\label{tc.pion}
\end{figure}
In \Ref{karsch}, the authors parametrized the $m_\pi$-dependence of $T_c$ 
in the full lattice QCD with $N_f=2$ as
\begin{equation}
	\left( {T_c \over \sqrt{\sigma}} \right)(m_{\pi})
=
	0.40(1)
+
	0.039(4)\left( {m_{\pi} \over \sqrt{\sigma}} \right),
\label{parameterize}
\end{equation}
where  $\sigma$ denotes  the string  tension.  
Strictly speaking, the phase  transition becomes just a cross-over for
intermediate values  of $m_{\pi}$. Hence, in  \Ref{karsch}, the pseudo
critical temperature is adopted as $T_c$, which is determined from the
peak positions of the susceptibilities of the Polyakov loop and so on.

For  the  $N_f=2$ case,  \Fig{tc.pion}  shows  the  estimate of  $T_c$
plotted  against $m_{\pi}$  based on  \Eq{fmt} in  the  closed packing
model. The dashed curve  denotes $T_c$ retaining only the contribution
from pions with $R_\pi \simeq 1$ fm.
The  solid curve denotes  the result  in the  case including  also the
contributions  from the low-lying  vector mesons,  such as  $\rho$ and
$\omega$, which are  the next lightest particles in  $N_f=2$ full QCD.
Here, we have used $\lambda_{\rho}=3\times 3=9$, $m_{\rho} = 770$ MeV,
$R_{\rho}\simeq  1$ fm,  $\lambda_{\omega}= 3$,  $m_{\omega}=783$ MeV,
$R_{\omega}\simeq   1$   fm   as   inputs,  which   are   treated   as
$m_{\pi}$-independent   constants.   These   vector  mesons   give  an
additional contribution  to $r_V(T)$ as $\delta r_V(T)  = 9 R_{\rho}^3
T^3 f(m_{\rho}/T) + 3 R_{\omega}^3 T^3 f(m_{\omega}/T)$.
The triangle and  the square in \Fig{tc.pion} denote  the lattice data
taken from Refs.\cite{karsch,luetgemeier}.
The $m_{\pi}$-dependence of the critical temperature $T_c$ is improved
after the inclusion  of the low-lying vector mesons,  i.e., $\rho$ and
$\omega$.
In spite of  the naively mindedness of the  closed packing model, this
statistical analysis seems to reproduce the lattice data, which may be
surprising.

\begin{figure}
\resizebox{0.48\textwidth}{!}{
  \includegraphics{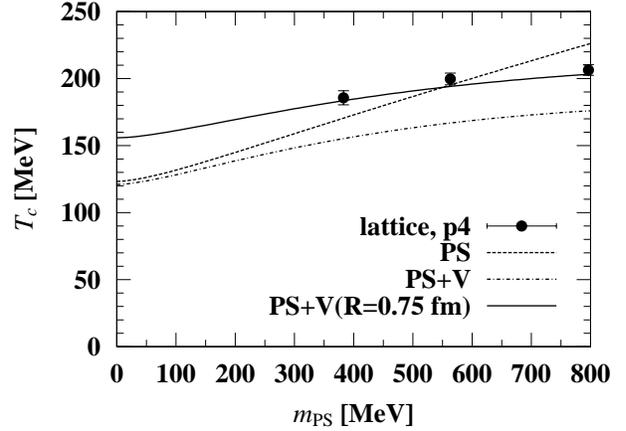}
}
\caption{The   critical   temperature   $T_c$  plotted   against   the
pseudo-scalar  meson  mass $m_{\rm{PS}}$  in  the SU(3)$_f$  symmetric
case.   The  dashed  curve  denotes  the  result  retaining  only  the
contribution of the pseudo-scalar mesons. The dot-dashed and the solid
curves  denote the  results  including also  the  contribution of  the
flavor-octet  vector  mesons  with  $R   =  1$  fm  and  $R=0.75$  fm,
respectively.   The  circle  denotes   the  lattice  data  taken  from
\Ref{karsch}  obtained  with   the  improved  staggered  (p4)  fermion
action.}
\label{tc.pseudo-scalar}
\end{figure}
We  next consider  the idealized  SU(3)$_f$ symmetric  case.   In this
case, the  pseudo-scalar (PS) octet  mesons, such as pions,  kaons and
$\eta_8$, are the  lightest and possess the same  mass $m_{\rm PS}$ in
common  with the  degeneracy  as $\lambda_{\rm{PS}}=8$.   We plot,  in
\Fig{tc.pseudo-scalar},   the  critical   temperature   $T_c$  against
$m_{\rm{PS}}$.   The dashed  curve  denotes $T_c$  retaining only  the
contribution from PS-mesons with $R_{\rm PS} \simeq 1$ fm, which leads
to $r_V(T)  = 8  R_{\rm{PS}}^3 T^3 f(m_{\rm{PS}}/T)$.   The dot-dashed
curve in \Fig{tc.pseudo-scalar} denotes the results including also the
contributions from  the octet  vector mesons with  $\lambda_{\rm{V}} =
8\times 3=24$, $m_{\rm{V}}\simeq 770$ MeV, $R_{\rm{V}}\simeq 1$ fm.
(Inclusion  of the  flavor-singlet vector  meson does  not  change the
result so much.)
These  vector mesons give  an additional  contribution to  $r_V(T)$ as
$\delta r_V(T)  = 24  R_{\rm{V}}^3 T^3 f(m_{\rm  V}/T)$.
The circle  in \Fig{tc.pseudo-scalar}  denotes the lattice  data taken
from \Ref{karsch}.
We  see that both  the dashed  and the  dot-dashed curves  are roughly
consistent with the lattice QCD results.
We  note  that  the  small  deviation almost  disappears  by  slightly
adjusting the bag  size as $R_{\rm{PS}} = R_{\rm{V}}  =0.75$ fm, as is
shown in \Fig{tc.pseudo-scalar} with the solid curve.

Thus, the simple statistical  approach with the closed packing picture
is  seen to  reproduce the  recent lattice  QCD data  on  the critical
temperature  $T_c$
and its $m_{\rm PS}$-dependence in  the full QCD phase transition both
for $N_f=2$ and $N_f=3$,
which would  suggest that  some of essential  natures of the  full QCD
phase transition could be governed by low-lying hadrons.
Considering its naivete and its simply minded nature, this coincidence
may be surprising.
It  would be  interesting to  refine  this approach  by including  the
interaction among hadrons.
However, we emphasize that the application of the closed packing model
to full QCD  is not the final aim of this  paper.
Rather, our aim is to  demonstrate the mysterious mismatch between the
critical temperature and the mass of the elementary excitation mode in
the quenched QCD phase transition in the next section.

\section{A mystery in the quenched QCD phase transition}
\label{sec.quench.qcd}

In this  section, we come  to the main  point of the paper,  i.e., the
crucial mismatch  of the critical temperature $T_c$  and the low-lying
elementary-excitation mode in the quenched QCD phase transition.
Focusing  on the  large difference  between the  lowest-lying glueball
mass  $m_{\rm G}  \simeq  1.5-1.7$ GeV  and  $T_c \simeq  280$ MeV  in
quenched QCD, we consider its physical implications.
To make the argument more  quantitative, we combine the recent lattice
QCD  data  with the  statistical  approach,  and  attempt to  make  an
estimate of the critical temperature $T_c$.

In quenched QCD, due to  the color confinement, only the color-singlet
modes can appear  as physical excitations.  Since all  of them consist
of color singlet combinations of gluons, they are called as glueballs.
The  mass spectrum  of the  glueballs  is known  through the  quenched
lattice  QCD  calculations  \cite{morningstar,weingarten,teper}.   The
lightest physical excitation is  the $0^{++}$ glueball with $m_{\rm G}
= 1.5-1.7$  GeV.
Being the  lightest physical excitation, the  lowest $0^{++}$ glueball
is  expected to  play the  key  role in  describing the  thermodynamic
properties of quenched QCD in  the confinement phase.  Hence, we first
take into account only the $0^{++}$ glueball.
As the mass  $m_{\rm G(S)}$ and the size $R_{\rm  G(S)}$ of the lowest
$0^{++}$ glueball, we adopt the  recent lattice QCD results as $m_{\rm
G(S)}  = 1730$  MeV \cite{morningstar}  and  $R_{\rm G(S)}  = 0.4$  fm
\cite{ishii,ishii.full.paper}.  We use  these values together with the
degeneracy  $\lambda_{\rm  G(S)}=1$   as  inputs.   Then  the  spatial
occupation  ratio  is  given   by  $r_V(T)  =  R_{\rm{G(S)}}^3$  $T^3$
$f(m_{\rm G(S)}/T)$, and the critical temperature is estimated as $T_c
\simeq 827$ MeV from  \Eq{critical.temperature}.  This estimate is too
much  larger  than   the  quenched  lattice  QCD  result   as  $T_c  =
0.629(3)\sqrt{\sigma}   \simeq   280$   MeV  in   \Ref{karsch2}   with
$\sqrt{\sigma} = 450$  MeV.  (In other words, only  a tiny fraction of
the  space region  can be  covered by  the thermally  excited  bags as
$r_V(T) = 0.0021$ at $T=280$  MeV.)  This large discrepancy would be a
serious problem in the quenched QCD phase transition.

To seek  the solution,  we examine the  possibilities of  the polemass
reduction and the  thermal swelling of the glueball  at $T\simeq T_c$,
which  are  suggested  in  \Ref{ichie}.
\begin{figure}
\resizebox{0.48\textwidth}{!}{
  \includegraphics{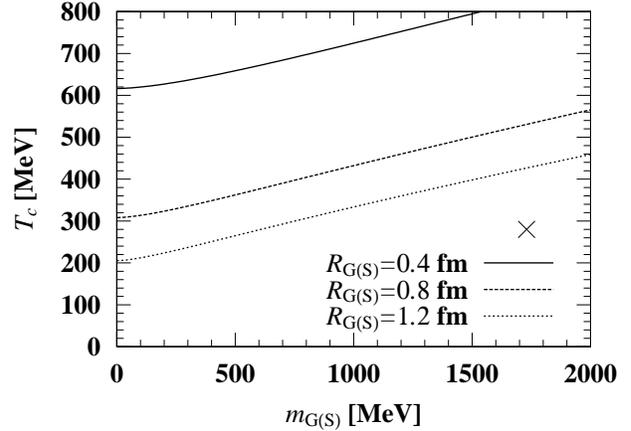}
}
\caption{The  critical temperature  $T_c$ plotted  against  the lowest
$0^{++}$  glueball mass $m_{\rm{G(S)}}$  in the  statistical approach.
The solid line denotes $T_c$ for the glueball size $R_{\rm{G(S)}}=0.4$
fm,     which     is     taken     from     the     recent     lattice
QCD\cite{ishii,ishii.full.paper}.  The  cross ($\times$) indicates the
quenched  lattice  QCD  results,   i.e.,  $T_c=280$  MeV  and  $m_{\rm
G(S)}=1730$ MeV.  We add  the cases with $R_{\rm{G(S)}}=0.8$ and $1.2$
fm by the dashed and dotted lines, respectively.  }
\label{tc-mG}
\end{figure}
We  first  consider  the  possibility  of the  polemass  reduction  at
$T\simeq  T_c$.  To  this end,  we  show in  \Fig{tc-mG} the  critical
temperature $T_c$ plotted against  the lightest $0^{++}$ glueball mass
$m_{\rm{G(S)}}$  in the  closed packing  model for  the  glueball size
$R_{\rm{G(S)}} = 0.4$ fm with the solid curve.
The  cross  ($\times$) indicates  the  quenched  lattice QCD  results,
$T_c=280$ MeV and $m_{\rm G}=1730$ MeV.
We see  that, for the problem  to be settled,  the polemass reduction
must be as significant as $m_{\rm{G(S)}}(T_c) \alt 500$ MeV.
However,     in     the     recent    lattice     QCD     calculations
\cite{ishii,ishii.full.paper}, it  has been reported  that the thermal
$0^{++}$ glueball persists to hold a rather large polemass as $m_{\rm
G(S)}(T \simeq T_c) \simeq 1250$ MeV. 
Hence, we examine another possibility, i.e., the thermal swelling.  In
\Fig{tc-mG}, we also include $T_c$ corresponding to different sizes of
the glueball as $R_{\rm G(S)} = 0.8, 1.2$ fm, which are denoted by the
dashed and dotted curves, respectively.
We see that to reproduce $T_c \simeq 280$ MeV, the thermal swelling of
the glueball must be quite significant at $T\simeq T_c$.
The   explicit  calculation   leads  to   an  abnormally   large  size
$R_{\rm{G(S)}} \simeq 3.1$ fm.
However, such a drastic thermal swelling of the lowest scalar glueball
was     rejected    by    the     recent    lattice     QCD    studies
\cite{ishii.full.paper,ishii}, which states  that the thermal glueball
size is almost unchanged even near $T_c$.

We seek for  another possibility by including the  contribution of the
excited-state glueballs.
In addition  to the lowest $0^{++}$ glueball  with $m_{\rm G(S)}\simeq
1730$ MeV  and $R_{\rm G(S)} \simeq  0.4$ fm, we  consider the thermal
contribution  from the  lowest $2^{++}$  glueball, which  is  the next
lightest hadron in quenched QCD.
We  take  $\lambda_{\rm{G(T)}}=5$,  $m_{\rm{G(T)}}  \simeq  2400$  MeV
\cite{morningstar}, $R_{\rm{G(T)}}\simeq 1$ fm. (Here, we assume it to
have a typical hadron size.)
Then, the  spatial occupation ratio  receives a correction  as $\delta
r_V(T) = 5  R_{\rm{G(T)}}^3 T^3f(m_{\rm{G(T)}}/T)$, and the correction
amounts to $\delta r_V(T)$ $=$ $0.0223$ at $T=280$ MeV.  The resulting
critical temperature is given as  $T_c \simeq 432$ MeV, which is still
too large.
We   note   that  the   realistic   glueball   size   would  be   more
compact. However, if so, its contribution becomes more negligible.
Besides these  two low-lying  glueballs, the following  excited states
are predicted in \Ref{morningstar} as $0^{-+}(2590)$, $0^{*++}(2670)$,
$1^{+-}(2940)$,   $2^{-+}(3100)$,   $3^{+-}(3550)$,   $0^{*-+}(3640)$,
$3^{++}(3690)$,   $1^{--}(3850)$,   $2^{*-+}(3890)$,   $2^{--}(3930)$,
$3^{--}(4130)$, $2^{+-}(4140)$, $0^{+-}(4740)$.
We include  all the contribution  from these excited  states, assuming
the  unknown glueball  size as  a typical  hadron size,  i.e., $R_{\rm
G}\simeq 1$  fm.  The  correction by these  excited states  amounts to
only  $\delta r_V(T)  = 0.0113$  at $T=  280$ MeV,  and  the resulting
critical temperature is  estimated as $T_c = 395$  MeV, which is again
too large.
Note that  we have used  $R_{\rm G}\simeq 1$  fm for the radii  of the
excited  glueballs,  since we  adopt  it  as  a typical  hadron  size.
However,  even  if we  adopt  $R_{\rm G}\simeq  2$  fm,  which may  be
considered well beyond the upper bound for a single hadron radius, the
total contribution to  $r_V(T)$ is $\delta r_V(T) =  0.269$ at $T=280$
MeV, which is still insufficient.

One  may  argue that  our  statistical  argument  can be  improved  by
considering  the  self-interaction of  the  glueball  and the  Lorentz
contraction effects.  However, unlike in full QCD, the glueball system
at $T\simeq T_c$ is so dilute that ideal-gas statistical treatment can
be applied. (We will come back  to this point in the summary.)  On the
other  hand, due to  the large  difference of  $T_c$ and  the glueball
masses,  the   effect  of  the  Lorentz   contraction  is  negligible.
Furthermore, the Lorentz contraction  effect drives the discrepancy in
the unwanted direction.

To recapitulate  this section, we  have observed that  the statistical
approach leads  to a  terrible overestimate of  $T_c$ in  quenched QCD
even after  the so many  improvement have been attempted.   The direct
cause  of  this failure  is  the  extremely  small statistical  factor
$e^{-m_{\rm{G(S)}}/T_c}$ $\simeq$  $0.0021$, which strongly suppresses
the  excitations of  the glueballs.   As a  consequence, only  a small
fraction of the  space region can be covered  by the thermally excited
bags of  glueballs even at  $T=280$ MeV, and $T_c$  becomes abnormally
large.

\section{Summary and discussions 
--- What is the trigger or the driving force of the QCD phase transition ?}
\label{sec.discussion}

We  have considered  the large  difference between  the  glueball mass
$m_{\rm  G} \simeq 1500-1700$  MeV and  the critical  temperature $T_c
\simeq 260-280$ MeV in the quenched QCD.
As a consequence of this large difference, the thermal excitation of a
single  glueball   is  suppressed  by  a   strong  statistical  factor
$e^{-m_{\rm G}/T_c} \simeq 0.00207$.   We have considered its physical
implications and argue  the abnormal nature of the  quenched QCD phase
transition.
To  appreciate  how abnormal  it  is,  we  have used  the  statistical
argument  with the  bag-model  picture of  hadrons,  i.e., the  closed
packing model.
We  have  derived  a  phenomenological  relation  among  the  critical
temperature $T_c$,  the mass  and the size  of the  low-lying hadrons.
We  have demonstrated  that  with slight  modifications  of the  model
parameters,   the    closed   packing   model    can   reproduce   the
$m_{\pi}$-dependence of $T_c$ obtained by full lattice QCD for $N_f=2$
and $3$  cases, suggesting that  some of the essential  ingredients of
the full QCD phase transition is governed by the lowlying hadrons.

Unlike full QCD, we have  found that the statistical approach terribly
overestimates  the critical  temperature as  $T_c \simeq  827$  MeV in
quenched QCD,  and that only  a tiny fraction  of the space  region is
covered  by  the thermally  excited  bags  of  glueballs at  $T\simeq$
280MeV.
We  have considered  the possibility  of the  thermal swelling  of the
glueball  size and  the polemass  reduction of  the glueball  near the
critical temperature.  However, both of these two have not provided us
with the solution on the large discrepancy.
Even though one includes all  the contributions  from the  15 low-lying
glueballs   up  to   5   GeV  predicted   in   quenched  lattice   QCD
\cite{morningstar},  the discrepancy  remains to  be still  large.  In
other words,  the number of  the thermally-excited glueballs  is still
too small  at $T \simeq$ 280MeV,  even after so  many glueball excited
states are taken into account.
The direct origin of this discrepancy is the strong suppression of the
thermal excitation of glueballs due to the extremely small statistical
factor as $e^{-m_{\rm{G(S)}}/T} = 0.00207$ at $T=280$ MeV even for the
lightest  glueball, which  leads  to the  insufficient  amount of  the
covered space by the thermally excited bags of glueballs.
We remark  that the  crucial role  is played by  the smallness  of the
statistical  factor $e^{-m_{\rm  G(S)}/T}$, which  has  rather general
nature.   Hence, through  this  failure, the  closed packing  argument
suggests a general tendency that,  if based on the lowlying excitation
modes,  the natural critical  temperature $T_c$  tends to  become much
larger in quenched QCD.
Actually, such a tendency is also found in the field-theoretical model
such  as the  dual Ginzburg-Landau  theory \cite{ichie},  which  is an
effective  theory of  color  confinement based  on  the dual  Meissner
effect.
It is remarkable that the  essential origin of this large deviation of
$T_c$  can  be understood  in  our  naive  statistical approach  in  a
simplified and idealized manner.

Although  several  arguments given  so  far  have  been based  on  the
bag-model  picture, this problem  itself has  a quite  general nature,
which can go beyond the reliability of the model framework.
Finally, we reformulate  this problem  in a  model-independent general
manner by  considering the inter-particle distance $l$  instead of the
bag size $R$.
Note that the reliability of the statistical argument becomes improved
in the  dilute glueball gas limit.   Now, taking into  account all the
low-lying 15 glueball modes up  to 5 GeV predicted in quenched lattice
QCD \cite{morningstar},  we calculate the  inter-particle distance $l$
of  the  glueballs based on only the statistical argument.
At $T=280$  MeV, the inter-particle distance is  estimated as $l\simeq
5$ fm.
It  follows that  the deconfinement  phase transition  takes  place at
$T\simeq$  280MeV,  where  the  glueball density is rather small as $\rho  =
1/\{\frac{4\pi}{3}(2.5{\rm  fm})^3\} \simeq 1/(4.0{\rm  fm})^3$.
Since the theoretical  estimates of  the  glueball size  are
rather      small     as      $R_{\rm{G(S)}}     \alt      0.4$     fm
\cite{ishii,ishii.full.paper,schafer}, the glueball system can be regarded  
to be dilute enough at this density just below $T_c$.
Furthermore,  since  the  long-range  interaction among  glueballs  is
mediated by the virtual one-glueball exchange process in quenched QCD,
the interactions  among glueballs are  exponentially suppressed beyond
its Compton length $1/m_{\rm{G(S)}} = 0.112$ fm. Therefore, one cannot
expect   the   strong   long-range   interaction  acting   among   the
spatially-separated thermal glueballs, and the dilute gas treatment of
the glueballs  is considered to be  valid at least  in the confinement
phase in quenched QCD.

Now, it is  quite difficult to imagine how such  a too rare excitation
of thermal  glueballs can  lead to the  phase transition.
We are thus arrive at the  mystery.
What is really  the trigger or the driving  force of the deconfinement
phase transition  in quenched QCD ?
Actually, also  in full QCD, the  trigger of the  QCD phase transition
would be  the interesting quenstion, but  not so many  have been known
yet.
In this sense,  the careful consideration on this  problem may provide
us with a key to find out the trigger of the QCD phase transition.

\begin{acknowledgement}
H.~S. is supported by Grant for Scientific Research (No.12640274) from
Ministry    of   Education,    Culture,   Science    and   Technology,
Japan.
\end{acknowledgement}

\end{document}